\documentclass[runningheads]{llncs}

\usepackage{graphicx}
\usepackage{amsmath}

\begin{document}
\title{Automated Lesion Detection by Regressing Intensity-Based Distance with a Neural Network} 
\titlerunning{Automated Lesion Detection by Regressing Intensity-Based Distance}

\author{Kimberlin M.H. van Wijnen\thanks{Both authors contributed equally to this work (\email{k.vanwijnen@erasmusmc.nl}, \email{f.dubost@erasmusmc.nl}).} \inst{1} \and 
Florian Dubost$^{\star}$\inst{1} \and  
Pinar Yilmaz\inst{2} \and 
M. Arfan Ikram\inst{2,3} \and 
Wiro J. Niessen\inst{1,4} \and 
Hieab Adams\inst{2} \and 
Meike W. Vernooij\inst{2} \and 
Marleen de Bruijne\inst{1,5}}  
\authorrunning{K.M.H. van Wijnen and F. Dubost et al.}

\institute{Biomedical Imaging Group Rotterdam, Department of Radiology \& Nuclear Medicine, Erasmus MC, Rotterdam, The Netherlands \and
Departments of Radiology \& Nuclear Medicine and of Epidemiology, Erasmus MC, Rotterdam, The Netherlands \and
Department of Neurology, Erasmus MC, Rotterdam, The Netherlands \and
Imaging Physics, Faculty of Applied Sciences, TU Delft, the Netherlands \and
Department of Computer Science, University of Copenhagen, Denmark}

\maketitle              
\begin{abstract}

Localization of focal vascular lesions on brain MRI is an important component of research on the etiology of neurological disorders. However, manual annotation of lesions can be challenging, time-consuming and subject to observer bias. Automated detection methods often need voxel-wise annotations for training. We propose a novel approach for automated lesion detection that can be trained on scans only annotated with a dot per lesion instead of a full segmentation. From the dot annotations and their corresponding intensity images we compute various distance maps (DMs), indicating the distance to a lesion based on spatial distance, intensity distance, or both. 
We train a fully convolutional neural network (FCN) to predict these DMs for unseen intensity images. The local optima in the predicted DMs are expected to correspond to lesion locations. We show the potential of this approach to detect enlarged perivascular spaces in white matter on a large brain MRI dataset with an independent test set of 1000 scans. Our method matches the intra-rater performance of the expert rater that was computed on an independent set. We compare the different types of distance maps, showing that incorporating intensity information in the distance maps used to train an FCN greatly improves performance.

\keywords{Lesion Detection  \and Geodesic Distance \and Fully Convolutional Neural Network \and Dot Annotations \and Perivascular Spaces}

\end{abstract}

\setcounter{footnote}{0}

\section{Introduction}

Obtaining the location of focal vascular lesions on brain scans, such as white matter hyperintensities, lacunes, enlarged perivascular spaces or microbleeds is extremely useful for studying the association of these lesions with neurological disorders.
However the manual annotation of these lesions can be challenging, time-consuming and subject to observer bias due to the difficulty of distinguishing a specific type of lesion from other similarly appearing structures. An automated method for detecting lesions could improve reliability, generalization and speed of lesion detection, which could greatly advance neuropathology research. 

Various promising automated methods have been proposed to detect lesions.
Deep learning methods often provide the best accuracy, but depend on expensive manual annotations for training like voxel-wise segmentations \cite{Brosch2016,Ghafoorian2017Lacunes} or bounding boxes \cite{Dou2016} marking the lesions. This  hinders applicability of these techniques in practice. 

Annotating by placing a single dot per lesion instead is considerably more time-efficient, allowing to collect larger annotated datasets for training and evaluation.
In this paper we therefore propose a novel method for lesion detection that requires only dot annotations.
Dot annotations have been effectively used to train convolutional neural networks (CNNs) for other applications, such as cell detection in histology images \cite{Xie2018Proxim}, lacune detection in placental ultrasound \cite{Qi2018LacuneUS} and landmark detection in retinal images \cite{Meyer2018Retina}. An approach that has shown great promise is regression of a distance map (DM) that is computed from these dot annotations \cite{Meyer2018Retina,Qi2018LacuneUS,Xie2018Proxim}.
Contrary to many other deep learning detection methods that use a two-stage approach \cite{Dou2016}, this approach directly outputs predicted detections and is optimized in an end-to-end fashion.  

We use a similar approach for detecting lesions based on dot annotations. Previous distance regression approaches for detection  \cite{Meyer2018Retina,Xie2018Proxim} have used Euclidean distance. This is especially suited for the detection of circular objects such as cells. Brain lesions on the other hand often have a morphology that is complex and discriminative \cite{Boespflug2018a}. 

In this paper we investigate the effect of including intensity information in DMs for lesion detection. Intensity distance incorporates local image context enabling the DM to capture complicated morphologies. Voxels surrounding dot annotations which have similar intensity values (inside the lesions) will have a lower value in the DM than dissimilar voxels (outside the lesions). This could encourage the CNN to learn the characteristic morphology of the lesions and propose more accurate detections than when trained on a Euclidean distance map (EDM) that does not make this distinction.
We compare Euclidean distance, intensity distance, and geodesic distance that combines both Euclidean and intensity distances. For geodesic distance the image is seen as a curved surface defined by the spatial coordinates and one intensity coordinate, where the shortest path on the surface is the geodesic distance \cite{Toivanen1996}. 

In this paper we show that including image intensity information in the DM improves optimization of a CNN for detecting lesions in brain MRI. We compute DMs from the dot annotations and their corresponding intensity images. Subsequently we train a fully convolutional neural network (FCN) to predict these DMs for unseen intensity images. The local minimal distances in the predicted DMs correspond to the proposed detection candidates.

We show the potential of regressing intensity-based DMs for the detection of enlarged perivascular spaces (PVS). PVS burden has been associated with cerebral small vessel disease \cite{Charidimou2013}. As PVS follow the course of the vessel they surround, they appear as elongated structures on 3D brain MRI scans.
Several methods have been proposed to detect PVS. The majority of the proposed algorithms is however evaluated on a relatively small sets (less than 30 images) due to the need for voxel-wise annotations for testing (and training) \cite{Boespflug2018a,Lian2018a}. We train and validate on a set of 1202 MRI scans and test on a separate set of 1000 images.
As the centrum semiovale (CSO) is seen as the most difficult brain region for PVS detection and most clinically relevant, we focused on this brain region \cite{Ballerini2018}.

\section{Method}

We train an FCN to regress a DM for a given intensity image. Our approach requires MRI scans with dot annotations for training. The local optima in the predicted DMs are expected to correspond to lesion locations.
We compare geodesic distance maps (GDMs), EDMs and intensity distance maps (IDMs).

\subsection{Distance Transform}
\label{sec:gdm}

To compute DMs we use a distance transform, that requires a definition of the foreground -- in our case the set of dot annotations $\Phi$ -- and a gray-scale image $G(x)$ in the case of intensity and geodesic distances, with $x$ the position in the image. 
The distance map $DM(x)$ is defined by

\begin{equation}
\label{eqn:GDM}
DM(x) = min(\Lambda(\gamma), \gamma \in \Psi(x,\Phi))
\end{equation}

with $\Psi(x,\Phi)$ the set of possible paths $\gamma$ between a position $x$ in the image and the set of dot annotations $\Phi$. The length $\Lambda(\gamma)$ of the path $\gamma$ is

\begin{equation}
\label{eqn:path}
 \Lambda(\gamma) = \sum_{i=1}^{n-1} d (x_i,x_{i+1})
\end{equation}

with $n$ the number of voxels in the path $\gamma$ between a position $x$ and a dot annotation $x_{dot}\in \Phi$ and $d$ the distance measure.
The geodesic distance $d_{G}$ in a 2D gray-scale image between voxel $x_i$ and the next voxel in the path $x_{i+1}$, with intensities $G(x_i)$ and $G(x_{i+1})$ respectively, is defined by \cite{Toivanen1996} as

\begin{equation}
\label{eqn:dgeo}
 d_{G}(x_i, x_{i+1}) =  \sqrt{d_{I}\big(x_i,x_{i+1}\big)^2 + d_{E}\big(x_i, x_{i+1}\big)^2}
\end{equation}

with the intensity distance $d_{I}(x_i,x_{i+1}) = G(x_i) - G(x_{i+1})$ and the Euclidean distance $d_{E}(x_i,x_{i+1})$ which is 1 for $x_{i+1} \in N_4(x_i)$ (voxels connected horizontally and vertically) and $\sqrt{2}$ for $x_{i+1} \in N_8(x_i) \setminus N_4(x_i)$ (voxels connected diagonally). EDMs are consequently computed by setting $d_{I} = 0$ in equation \ref{eqn:dgeo}, while IDMs are computed by setting $d_{E} = 0$.
We approximate these DMs using the optimization algorithm \textit{iterative raster scan} described in \cite{Toivanen1996}. This approach is for computing DMs in 2D, though it can easily be extended to 3D.\footnote{Our code for computing 2D as well as 3D distance maps is available at \email{https://github.com/kimvwijnen/geodesic\_distance\_transform}}

The resulting $DM(x)$ is normalized by dividing by the maximum distance in the $DM(x)$ and inverted as this is convenient for implementation. Furthermore, we add a parameter $p$ to influence how steeply the distance decays. The final map $M_{p}(x)$ is calculated using

\begin{equation}
\label{eqn:LabIm}
M_{p}(x) = \Big( 1 - \frac{DM(x)}{max\big(DM(x)\big)} \Big)^p
\end{equation}

\subsection{Fully Convolutional Neural Network}

\begin{figure}[t]
\centering
\includegraphics[width=\textwidth]{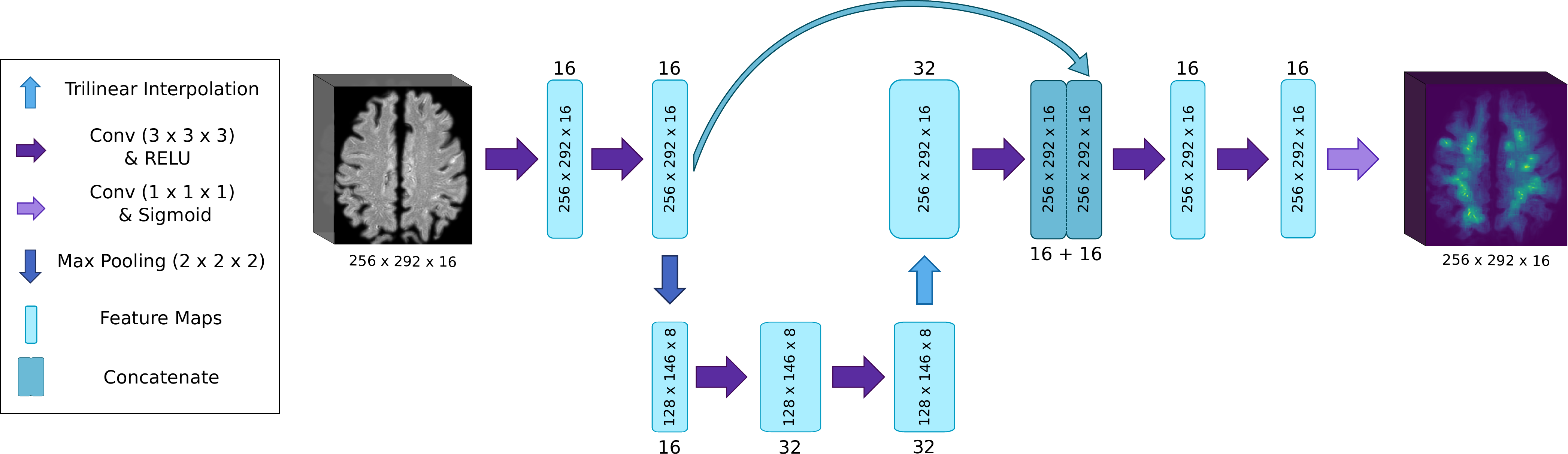}
\caption{Network architecture, on the left the input preprocessed brain scan is shown and the output predicted distance map is shown on the right} \label{fig:unet}
\end{figure}

We use an architecture similar to a shallow U-Net for our FCN shown in Figure \ref{fig:unet}, which was shown to work well for regressing the number of perivascular spaces in the basal ganglia \cite{Dubost2017,Ronneberger2015}. For optimization we use mean square error loss \mbox{$MSE =  \frac{1}{N}  \sum\limits_{x} \Big(\widehat{M}_{p}(x) - M_{p}(x)\Big)^2$}, with $\widehat{M}(x)$ the predicted map and $N$ the number of voxels in $M_{p}(x)$.

Non-maximum suppression is applied to the predicted distance map to detect local optima. We use a $5 \times 5$ maximum filter with a connectivity of 8. By thresholding the local optima the proposed detections are acquired.

\section{Experiments}
\subsection{Data}
\label{sec:data}
Our data set consists of 2202 T2-weighted MRI scans from the Rotterdam Scan Study.
All scans were from different individuals and were acquired on a 1.5 T MRI scanner.
The images have a size of $512 \times 512 \times 192$ with a voxel resolution of $0.49 \times 0.49 \times 0.8 \mbox{mm}^{3}$.
Further details on the image acquisition of this data are discussed by Ikram et al. \cite{Ikram2015}. 

The number of PVS in the axial slice 1 cm above the lateral ventricles is highly correlated with the total number of PVS in the CSO \cite{Adams2015}. The rater selected this specific slice and annotated it with dots indicating PVS between 1 - 3 mm in diameter in line with the guidelines described by Adams et al. \cite{adams2013rating}. 
The intra-rater performance was evaluated on a separate set of 40 MRI scans (see Table \ref{tab:faucs} and Figure \ref{fig:frocmse}).

\subsection{Preprocessing}
Images are preprocessed as proposed by \cite{Dubost2018AllReg}. 
We segment the CSO with the FreeSurfer multi-atlas segmentation algorithm \cite{Desikan2006} producing a binary mask that we smooth with a Gaussian kernel. The image are multiplied with the smoothed mask and cropped to a fixed size containing only the slices close to the annotated slice.
The resulting images are normalized to the range [0,1] by dividing by the maximum intensity in the image. 

Annotated dots were not always inside the PVS. To solve this problem, we shift the dots to the highest intensity value within the same connected component and within 3 voxels distance. The shifted dots were only used to compute the distance maps for the training and validation set. For evaluation of the detection performance, the original annotated dots were used.

\begin{figure}[b]
\centering
\includegraphics[width=\textwidth]{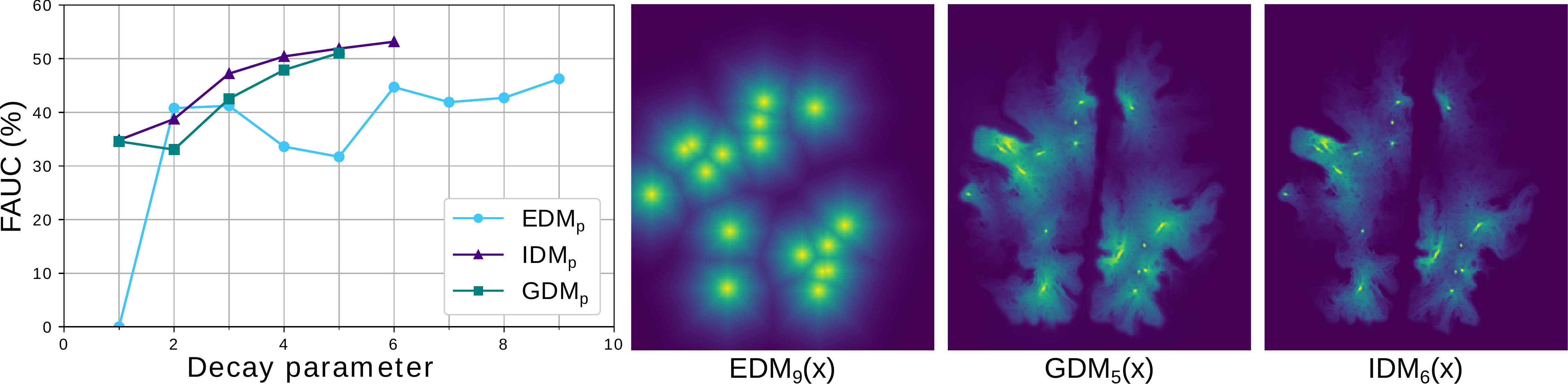}
\caption{Influence of decay parameter $p$ on detection performance on the validation set and the chosen distance maps} \label{fig:dms}
\end{figure}

\subsection{Experimental Setup}

Random sampling was used to split the 2202 scans into a set of 1202 for development of the method (1000 for training and 202 for validation) and a separate set of 1000 for testing. 
As only one slice per scan was annotated, DMs were computed in 2D and the loss was only evaluated for this slice. Non-maximum suppression and evaluation of detection performance was also only done on the slice that was annotated.

Weights for the convolutional layers were initialized by random sampling from a truncated normal distribution with zero mean and unit variance.
For optimization we use Adadelta and a batch of one due to memory limitations. We use on-the-fly augmentation for the training set. For every image a random rotation around the depth direction with a maximum of $20^\circ$ in both directions is applied combined with random flipping in horizontal and in vertical direction.
Methods were implemented in Python and Keras with Tensorflow as backend.

\subsection{Detection Performance}

The candidate detections of each method are compared to the expert annotations using the hungarian algorithm to find a one-to-one mapping between these sets. Only detections within a 6 voxel radius of the annotations were counted as true positive. We use 6 voxels as this is the maximum PVS diameter (corresponds to 3 mm \cite{adams2013rating}).

The detection performance is mainly evaluated with the Free-Response Operating Characteristic (FROC) curve and its area under the curve (FAUC) until 10 $FP_{avg}$, which is approximately twice the $FP_{avg}$ of the rater. The FAUC is calculated as the percentage of the highest possible area. We used bootstrapping to quantify the uncertainty, resulting in a mean FAUC and confidence interval based on 1000 sampled sets. Bootstrapping was performed by random sampling with replacement from the test set.

\begin{figure}[t]
\centering
\includegraphics[width=\textwidth]{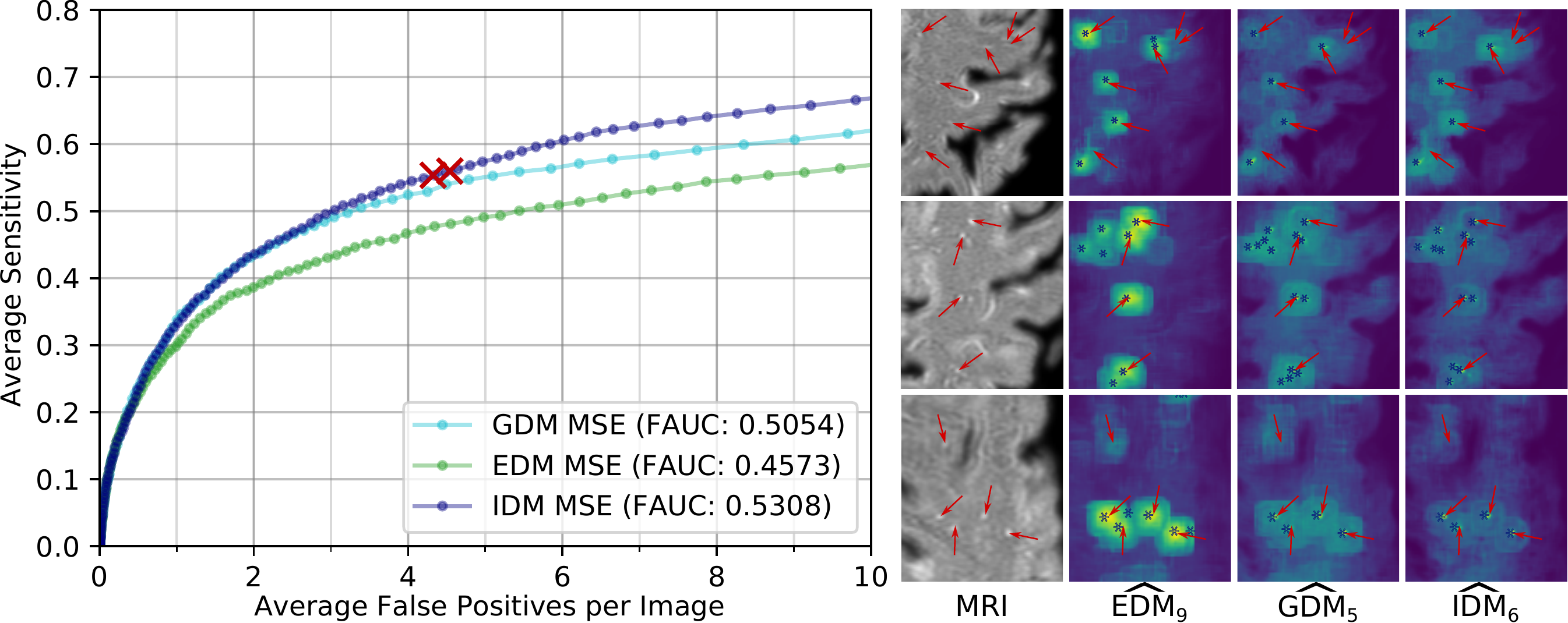}
\caption{FROC curves and crops of the output of the FCNs and their proposed detections (intra-rater performance indicated with red crosses, annotations with red arrows and predictions with blue stars)} \label{fig:frocmse}
\end{figure}

\subsection{Evaluation Approach}
\label{sec:eval}

We ran experiments varying the decay parameter $p$ (see Figure \ref{fig:dms}). For higher values of $p$ the FCN did not train, we expect because of label imbalance. Based on the FAUC on the validation set we set p to 5 for geodesic distance, to 6 for intensity distance and 9 for euclidean distance. During training, the model parameters were chosen as the ones minimizing the FAUC computed on the validation set. Only the best model per distance type ($GDM_{5}(x)$, $EDM_{9}(x)$, $IDM_{6}(x)$) was tested on the test set of 1000 scans. 

The operating point on the FROC was chosen per model as the threshold with a sensitivity on the validation set closest to the average intra-rater sensitivity. 
For $GDM_{5}(x)$ the threshold was chosen at 0.525, for $EDM_{9}(x)$ at 0.500 and for $IDM_{6}(x)$) at 0.495.
This threshold was used as the detection threshold during evaluation on the test set.

\subsection{Results}
Figure \ref{fig:frocmse} shows the FROC curves computed on the test set and examples of the output of the FCNs. Table \ref{tab:faucs} shows the corresponding FAUCs, the sensitivity and $FP_{avg}$ of the methods on the test set at the chosen thresholds (based on the validation set) and the average intra-rater performance.

\section{Discussion and Conclusion}

Our experiments indicate that incorporating image intensity information in a distance map used to train an FCN substantially improves performance of PVS detection. Results show that using GDMs and IDMs both result in a similar detection performance, with IDMs sometimes reaching higher performance than GDMs. This indicates that intensity difference is the most discriminative information, and that Euclidean distance could even be ignored.
Using higher values of the decay parameter also increases the PVS detection performance, and stabilizes the optimization. 

The FCN trained using IDMs reaches a sensitivity and $FP_{avg}$ similar to the intra-rater performance computed on a smaller independent set (Figure \ref{fig:frocmse}). 

We expect our method could perform well for detecting other types of focal vascular lesions in the brain. Using intensity information in the computation of DMs could help the detection lesions that either have a complex morphology, or can have substantial variation in their size, such as microbleeds, white matter hyperintensities or lacunes. 
Additionally, in this work we evaluate the intensity-based distance maps only for their performance in detecting PVS.  However, we observe that the PVS detections in the output maps of the FCNs trained on intensity-based distance maps (Figure \ref{fig:frocmse}) seem to approximate the PVS shape quite well. We therefore expect our approach might also work well for segmentation. 

\begin{table}[t]
\caption{PVS detection performance on the test set for the detection methods and the average intra-rater performance on a smaller independent set}\label{tab:faucs}
\centering
\begin{tabular}{l l l l}
\hline
					& FAUC \hspace{6em} & $FP_{avg}$ \hspace{2em} & Sensitivity \\
\hline	
\textbf{$EDM_{9}(x)$}  		& 45.761 ($\pm$  0.052) 	& 7.49
	& 	53.63	\\
\textbf{$GDM_{5}(x)$}			& 50.575 ($\pm$  0.050) 	& 5.10	&	55.26 	\\
\textbf{$IDM_{6}(x)$}   		& 53.078 ($\pm$  0.051)  	& 4.35
	&	55.35	\\
Average intra-rater  \hspace{2em}		& -  	& 4.43
	&	55.66	\\
\hline
\end{tabular}
\end{table}

\subsubsection{Acknowledgments}
This research is part of the research project Deep Learning for Medical Image Analysis (DLMedIA) with project number P15-26, funded by the Dutch Technology Foundation STW (part of the Netherlands Organisation for Scientific Research (NWO), which is partly funded by the Ministry of Economic Affairs), and with co-financing by Quantib. This research was also funded by the Netherlands Organisation for Health Research and Development (ZonMw), Project 104003005.
Part of this work was carried out on the Dutch national e-infrastructure with the support of SURF Cooperative and on a Quadro P6000 donated by the NVIDIA Corporation.

\bibliographystyle{splncs04}
\bibliography{paper2108}

\end{document}